# Application of machine learning to predict food processing level using Open Food Facts

Nalin Arora[1,3,4], Aviral Chauhan[2,3,4], Siddhant Rana[2,3,4], Mahansh Aditya[3,4,5], Sumit Bhagat[2,3,4], Aditya Kumar[3,4,6], Akash Kumar[3,4,6], Akanksh Semar[3,4,7], Ayush Vikram Singh[2,3,4] and Ganesh Bagler[1,3,4,8],*

[1]Department of Computational Biology
[2]Department of Computer Science
[3]Infosys Center for Artificial Intelligence
[4]Center of Excellence in Healthcare
[5]Human Centered Design
[6]Department of Mathematics
[7]Department of Electronics & Communications Engineering

Indraprastha Institute of Information Technology Delhi (IIIT-Delhi), New Delhi 110020 India.

[8]Foodoscope Technologies Pvt Ltd, New Delhi, India.

**\*Corresponding Author:** Ganesh Bagler

Infosys Center for Artificial Intelligence
Department of Computational Biology
A-305 (R&D Block), Indraprastha Institute of Information Technology Delhi (IIIT-Delhi)
Okhla Phase III, New Delhi, India, 110020
Phone: 91-11-26907-443
Email: bagler@iiitd.ac.in

## ABSTRACT

Ultra-processed foods are increasingly linked to health issues like obesity, cardiovascular disease, type 2 diabetes, and mental health disorders due to poor nutritional quality. This first-of-its-kind study at such a scale uses machine learning to classify food processing levels (NOVA) based on the Open Food Facts dataset of over 900,000 products. Models including LightGBM, Random Forest, and CatBoost were trained on nutrient concentration data. LightGBM performed best, achieving 80–85% accuracy across different nutrient panels and effectively distinguishing minimally from ultra-processed foods. Exploratory analysis revealed strong associations between higher NOVA classes and lower Nutri-Scores, indicating poorer nutritional quality. Products in NOVA 3 and 4 also had higher carbon footprints and lower Eco-Scores, suggesting greater environmental impact. Allergen analysis identified gluten and milk as common in ultra-processed items, posing risks to sensitive individuals. Categories like Cakes and Snacks were dominant in higher NOVA classes, which also had more additives, highlighting the role of ingredient modification. This study, leveraging the largest dataset of NOVA-labeled products, emphasizes the health, environmental, and allergenic implications of food processing and showcases machine learning's value in scalable classification. A user-friendly web tool is available for NOVA prediction using nutrient data: https://cosylab.iiitd.edu.in/foodlabel/.

## Keywords

# 1. Introduction

Historically defined by the NOVA food classification system, ultra-processed foods (UPFs) include a wide variety of ready-to-eat products—ranging from packaged snacks and carbonated drinks to ready-to-eat meals (Monteiro et al., n.d.-a). These products are predominantly composed of food derivatives, incorporating flavors, colors, texturizers, and other additives, with minimal inclusion of whole foods (Monteiro et al., 2019a). Recently, UPFs have been recognized as industrially processed products formulated with multiple ingredients, including salts, sugars, oils, fats, and various additives, distinguishing them from traditional culinary preparations (Gibney, 2018). The excessive consumption of UPFs has led to the rise in health disorders associated with their intake (Popkin, 2006).

Several studies have identified potential associations between ultra-processed food consumption and various health conditions, including obesity, high blood pressure, and metabolic syndrome, though causal relationships remain to be established (Santos et al., 2020). While studies indicate correlations between UPF intake and non-communicable diseases such as type 2 diabetes and cardiovascular diseases (Chen et al., 2020), and cardiovascular events (Qu et al., 2024), the varying applications of NOVA criteria combined with the different ways researchers have categorized UPFs, make it challenging to draw definitive conclusions. Studies have suggested connections between UPF consumption and increased risk of several cancers, including colorectal, breast, and pancreatic cancers (Isaksen & Dankel, 2023), and high-dose UPF intake has been linked to chronic kidney disease (Xiao et al., 2024). Similarly, observations indicate that individuals consuming higher amounts of UPFs show increased likelihood of experiencing mild depression, mental health issues, and anxiety, though more research is needed to understand these relationships (Hecht et al., 2022).

The reported consumption patterns of UPFs show considerable variation globally, with higher intake in countries like the United States (60% of total caloric intake) and the United Kingdom (58%), compared to lower consumption in Italy (10%) and South Korea (25%) (Lane et al., 2024; Mertens et al., 2022). The energy share from UPFs varies across Europe, ranging from 14% in Italy to 44% in Sweden, with fine bakery wares and soft drinks being primary contributors (Mertens et al., 2022). The Nutri-Score system, which provides a standardized method for evaluating the nutritional quality of food products, indicates that UPFs generally score poorly due to their high levels of sugars, fats, and additives, while lacking essential nutrients (Lane et al., 2024). While this single letter grading approach may oversimplify the multifaceted nature of food quality and nutritional value, it remains an essential criterion. Studies have shown that the Nutri-Score significantly improves the nutritional quality of purchasing intentions, encouraging healthier food choices, particularly among individuals suffering from cardiometabolic chronic diseases (Egnell et al., 2022). The Eco-Score system, which evaluates the environmental impact of food products through a comprehensive life cycle assessment, reveals a complex relationship between processing levels and environmental impact, with UPFs often showing lower greenhouse gas emissions compared to minimally processed foods (MPFs) (Aceves-Martins et al., 2022). Research has demonstrated that the Eco-Score label effectively influences consumer behavior, reducing the selection of environmentally harmful products by 13% while increasing the choice of sustainable options by 16% (Taillie et al., 2024).

In addition to these concerns, UPFs are characterized by their high content of additives and an increased likelihood of containing allergens. Recent studies have demonstrated that ultra-processed foods pose challenges regarding allergen content, as their complex manufacturing processes and multiple ingredients increase the risk of both intentional inclusion and cross-contamination of allergenic substances (Katidi et al., 2023). Further, UPFs contain more allergenic ingredients than less processed foods, with an average of 1.3 allergenic ingredients per UPF compared to 0.4 in MPFs (Amaraggi et al., 2024). Milk, eggs, fish, crustaceans, tree nuts, peanuts, wheat, and soybeans are thought to account for 90% of food allergy reactions (Messina & Venter, 2020). In Europe, 14 allergens are recognized as the most common and potent causes of food allergies and intolerances, namely, cereals containing gluten, milk, eggs, nuts, peanuts, soybeans, fish, crustaceans, mollusks, celery, lupin, sesame, mustard, and sulfites (*Allergens in Food: Scientific Advice Updated | EFSA*, n.d.).

Several food processing classification systems exist, each with its own categorization procedure. These include systems developed by the International Food Information Council (Eicher-Miller et al., 2012), the International Agency for Research on Cancer based on methods from the European Prospective Investigation into Cancer and Nutrition study (Chajès et al., 2011; Moubarac et al., 2014), the National Institute of Public Health in Mexico, and the International Food Policy Research Institute in Guatemala (Asfaw, 2011; Moubarac et al., 2014). Among these, the NOVA system, developed in Brazil, stands out as the most distinguished (Monteiro et al., n.d.-b). NOVA offers a universally relevant framework due to its precision, coherence, and comprehensiveness in categorizing foods based on their level, scope, and intent of industrial processing. Its versatile nature allows for global adaptation (Moubarac et al., 2014), while other systems like the UNC system developed by the University of North Carolina have been specifically designed to address the classification needs of U.S. supermarket products (Bleiweiss-Sande et al., 2019; Poti et al., 2015). Despite its limitations, including low inter-rater reliability (K = 0.32-0.34) among food experts and having classified only 35% of U.S. Department of Agriculture cataloged foods, NOVA remains the most widely recognized and utilized system globally (Braesco et al., 2022). These limitations, rather than invalidating NOVA's use, underscore the critical need for automated classification systems, the lacuna that this study aims to address.

Efforts to automate classification include studies by Menichetti et al., Hu et al. and Arora et al., employing machine learning (ML) and natural language processing (NLP) techniques (Arora et al., 2025; Hu et al., 2023; Menichetti et al., 2023). The studies utilized the Food and Nutrient Database for Dietary Studies (FNDDS) dataset (2009-10) along with NOVA labels from Steele et al., focusing exclusively on 2,970 food products, as these were the only products for which NOVA labels were available (*FNDDS DOWNLOAD DATABASES : USDA ARS*, n.d.; Monteiro et al., 2019b). Recent advances have demonstrated remarkable success in this domain - Arora et al. achieved Matthews Correlation Coefficient (MCC) of 0.86 using ML-based approaches and an even higher MCC of 0.9 using NLP-based transformer models, validating the potential of automated classification systems (Arora et al., 2025). Furthermore, the FoodProX algorithm achieved consistently high area under curve values (0.9804 for NOVA 1, 0.9632 for NOVA 2, 0.9696 for NOVA 3, and 0.9789 for NOVA 4) using nutrient profiles as predictive indicators (Menichetti et al., 2023). Additionally, transformer-based language models have shown exceptional accuracy in predicting food processing levels, with studies achieving F1 scores of up to 0.979 (Hu et al., 2023). In contrast, the present study leverages approximately 800,000 products from Open Food Facts (OFF) (*Https://In.Openfoodfacts.Org/*, n.d.). The previous studies lack methods to address missing data. To address this issue, we have implemented two strategies: Imputation of missing data with the mean, and the use of simple autoencoders (AEC) to fill-in the missing values, as they have shown to learn the representation of the data with missing values and generate plausible replacements (Pereira et al., 2020).

The present work develops a ML model to predict food processing levels based on nutrient features, utilizing the vast OFF database. While NOVA classification traditionally relies on processing methods and ingredients rather than nutrient content, our research as well as previous research demonstrate that nutrient profiles can serve as reliable predictive indicators of processing levels. Our strategic choice of nutrients as input features is grounded in three key principles: First, nutrient reporting follows consistent global regulations, ensuring standardized data across different regions and manufacturers. Second, unprocessed foods contain nutrients within physiologically constrained ranges determined by natural biochemistry, establishing reliable baseline patterns (Menichetti & Barabási, 2022). Third, industrial food processing creates systematic and reproducible alterations in nutrient concentrations, producing distinctive patterns that machine learning can detect (Menichetti et al., 2023). These nutrient alterations serve as quantifiable markers of processing level, allowing our model to accurately predict NOVA classifications without directly analyzing processing methods. By leveraging these nutrient-based patterns in the detailed information from thousands of food products worldwide, we created a model that accurately classifies food products according to the NOVA system. To support our primary objective, we conducted comprehensive exploratory data analysis to elucidate the relationships between various categorical variables (category, Eco-Score grade, Nutri-Score grade, and allergens) and NOVA classes. This analysis examines NOVA class distributions across geographic regions, food categories, and nutritional quality indicators, as well as investigates the correlation between allergen presence and food processing levels. Additionally, we analyze associations between continuous variables (carbon footprint, Nutri-Score, number of additives, and Eco-Score) and food processing levels, employing statistical methods and data visualization techniques to identify significant correlations and trends. We present the first comprehensive analysis of the OFF database, incorporating nutritional and environmental factors to predict food processing levels.

## 2. Materials and Methods

### 2.1 Dataset description and pre-processing

The nutrient data used in this study was obtained from OFF. OFF is an extensive, open-source database offering detailed nutritional information for various commonly consumed foods globally. We used the NOVA class parameter in the data to classify foods by their processing levels. Initially, we aligned the nutrients from OFF with those in the FNDDS (2009-10) data (*FNDDS DOWNLOAD DATABASES : USDA ARS*, n.d.), identifying 44 common nutrients. The 2009-10 FNDDS dataset was specifically chosen because it includes NOVA labels from Steele et al., allowing us to validate the models effectively. For further analyses, we selected 8 nutrients with missing values of 41% or less and eventually focused on 7 nutrients with missing values of 15% or less. Refer to Supplementary Table 1, to see the table for missing values and percentages of all nutrients.

For the first experiment, we used 875,075 food products. In subsequent experiments, the number of products decreased to 681,950 for the study involving 7 nutrients and 479,090 for the study with 8 nutrients. In the initial experiment, we imputed missing data using the mean and through AEC due to the high volume of missing values. For the later two experiments, we excluded food products with missing values.

**2.2 Exploratory Data Analysis**

We filtered the dataset for the top 50 most frequent categories within each NOVA class, creating a directed graph where nodes represent categories and edges show their co-occurrence. Using Louvain community detection, we grouped frequently co-occurring categories, coloring nodes by community, sizing them by degree, and filtering edges to reduce clutter. To visualize the distribution of allergens across NOVA classes, we first filtered the dataset to retain only food products containing NOVA class information and allergen data. Each allergen entry was split into individual records by separating comma-delimited lists. We created doughnut charts for each NOVA class, highlighting the top 10 allergens per class. Additionally, Nutri-Score and Eco-Score distributions were analyzed by filtering food items with valid grades (A-E) for both scoring systems. Contingency tables were created to compare the frequency of Nutri-Score and Eco-Score grades across NOVA classes, and stacked bar charts were used for visualization, with customized legends and labels for clarity. Finally, we visualized continuous features such as Carbon Footprint, Nutri-Score, and Eco-Score using boxplots to examine how their medians and other statistical measures vary across different NOVA classes (Krzywinski & Altman, 2014).

We employed a range of statistical tests to assess the relationships among NOVA classes and factors like Categories, Allergens, Nutri-Score, and Eco-Score. The Chi-Square test was used to evaluate the independence between categorical variables, with sample sizes of 794,163 for Categories, 431,854 for Allergens, 677,073 for Nutri-Score Grade, and 415,502 for Eco-Score Grade. Cramér's V measured the strength of association between these nominal data. We uses of the Chi-Square test due to its suitability for categorical variables. The alpha level for all tests was set at 0.05, and all tests were two-tailed. Additionally, we applied the Kruskal-Wallis test to compare differences in continuous features across NOVA classes, with sample sizes of 385 for Carbon Footprint and 677,073 for Nutri-Score and 415,502 for Eco-Score. We also assessed the correlation between variables to understand the strength and direction of their linear relationships. We use p-value as 0.05 throughout the course of the study.

**2.3 Autoencoder development**

AEC were developed using pytorch library in python. We initially filled the data with mean values and performed z-score normalization (Yun & Kim, 2024). The autoencoder architecture comprised fully connected layers in both the encoder and decoder, with adjustable dimensions and activation functions. Hyperparameters were optimized using Optuna (Akiba et al., 2019), and we focused on the following parameters: 'latent dimension', 'number of hidden layers', 'hidden layer dimensions', 'activation functions', 'learning rate', 'dropout probability', 'batch size', and 'weight decay.' The autoencoder was trained using the Adam optimizer and a mean squared error loss function for 50 trials of hyperparameter tuning. The final model was trained for 500 epochs using the best hyperparameters identified by Optuna (Akiba et al., 2019).

**2.4 Model development**

We focused on three tree-based ML models to predict food products' NOVA food processing levels: Random Forest (RF), LightGBM (LGBM), and CatBoost (CB). RF is an ensemble method that constructs multiple decision trees and combines their results to improve predictive performance (Breiman, 2001). It is known for the ability to handle high-dimensional data and robustness to outliers and noise. LGBM, conversely, is a gradient-boosting framework that utilizes tree-based algorithms to achieve high efficiency and accuracy (Ke et al., n.d.). It is designed to handle large-scale data and has been shown to outperform other boosting algorithms regarding training speed and memory usage. CB, another gradient boosting algorithm, is particularly effective in handling categorical features and overfitting (Prokhorenkova et al., 2017). It employs a combination of categorical feature encoding and ordered boosting to improve model performance.

To optimize the hyperparameters of these models, we employed RandomizedSearchCV from the scikit-learn library. This technique randomly searches over a specified parameter space, allowing for efficient exploration of various hyperparameter combinations (Bergstra et al., 2012). Using RandomizedSearchCV, we found the best set of hyperparameters for each model, thereby maximizing their predictive performance on the NOVA food processing level classification task.

For LGBM, we focused on tuning 'boosting_type', 'num_leaves', 'max_depth', 'learning_rate', 'n_estimators','subsample','colsample_bytree','reg_alpha','reg_lambda','min_split_gain','min_child_weight',and 'scale_pos_weight.' For RF, we tuned 'max_depth', 'n_estimators', 'max_features', 'min_samples_split', 'min_samples_leaf', and 'bootstrap.' Finally, we tuned 'iterations', 'depth', 'learning_rate', and 'l2_leaf_reg' for CB.

## 2.5 Handling class imbalance

To address the significant class imbalance in the dataset, three methods were applied: Synthetic Minority Oversampling Technique (SMOTE), which generates synthetic samples of the minority class to equalize class distribution using existing entries (Chawla et al., 2002); stratified k-fold cross-validation, which ensures that each fold maintains the same class proportions as the original dataset (Prusty et al., 2022); and a combination of SMOTE with stratified k-fold, which applies SMOTE within each fold to sustain balance throughout the training process.

## 2.6 Evaluation metrics

A range of evaluation metrics was used to assess model performance comprehensively. Accuracy, F1-Score, and MCC were the metrics employed, delivering a well-rounded view of each model's predictive performance by covering various aspects of classification accuracy, balance, and robustness. While MCC provides valuable insights into prediction balance, the primary focus was placed on Accuracy and F1-Score, as these metrics better capture the model's ability to handle the class imbalance inherent in the dataset.

## 2.7 Feature Importance

We used SHapley Additive exPlanations (SHAP) to interpret model predictions and identify key features influencing outcomes. SHAP values, derived from cooperative game theory, quantify each feature's contribution to predictions (Lundberg & Lee, 2017). This approach provided insights into feature importance and model behavior at both global and local levels.

## 3. Results
### 3.1 Open Food Facts data characterization

In Figure 1, the networks of food categories across NOVA classes illustrate foods' interconnectedness as their processing levels rise from NOVA 1 to NOVA 4. In NOVA 1, the graph is relatively sparse, with distinct clusters of MPFs such as 'Fruits,' 'Dairy products,' and 'Fermented foods,' reflecting the simplicity and limited connections between these categories. In NOVA 2, the network becomes more connected, with categories like 'Plant-based foods' and 'Cereals' showing more interactions, indicating a broader use of basic processed ingredients. NOVA 3 presents a denser graph with more complex interactions between categories like 'Snacks,' 'Biscuits,' and 'Sweetened beverages,' highlighting the shared use of ingredients in processed foods. Finally, NOVA 4 shows the most intricate and densely connected network, with highly processed categories like 'Cakes,' 'Sweets,' and 'Prepared meals' forming tightly knit communities, reflecting the dominance of ultra-processed foods in modern diets due to their shared industrial ingredients and extensive processing.

Based on the analysis of 270,088 food products, the prevalence of allergens across the four NOVA classes reveals some intriguing relations as shown in Figure 2. In NOVA 1, gluten dominates, representing 40% of the allergens, followed by nuts (17%), milk (15%), and peanuts (11%). NOVA 2 is overwhelmingly dominated by milk, which accounts for 71% of allergens, with smaller contributions from gluten (9%), nuts (6%), and peanuts (5%). NOVA 3 shows a more balanced distribution, with milk (32%), gluten (27%), and nuts (9%) being the top contributors. In NOVA 4, milk again leads at 34%, followed by gluten (25%) and nuts (14%). Across all classes, milk and gluten are the most prevalent allergens, with nuts and peanuts also making significant appearances.

Figure 3 visualizes the relationship between NOVA classes and Nutri-Score grades, revealing distinct patterns that highlight the connection between food processing levels and nutritional quality. Nutri-Score, a front-of-pack label, rates the nutritional quality of food products from A (healthiest) to E (least healthy) based on sugar, fat, salt content, and beneficial nutrients like fiber and protein. NOVA 1 stands out with a higher proportion of products in the healthier grades (A and B), reflecting better overall nutritional profiles. Specifically, only 4.98% and 3.30% of products in NOVA 1 fall into the lower grades D and E, with 4,053 and 2,685 items respectively, out of 81,298 items. NOVA 2 shows the lowest overall counts across all grades. However, it is notable that 36.5% and 15.22% of its products fall into grades D and E, with 10,537 and 4,394 items, respectively, out of a total of 28,865 items. NOVA 3 exhibits a more balanced distribution of Nutri-Score grades, with 29.35% and 10.74% of products in grades D and E, totaling 44,400 and 16,251 items, respectively, out of 151,229 items. NOVA 4 has the highest overall count, with a significant concentration in the lower Nutri-Score grades (D and E). 33.11% and 23.84% of its products fall into these grades, totaling 137,484 and 98,996 items, respectively, out of 415,181 items.

Figure 4 compares NOVA classes with Eco-Score grades, uncovering distinct patterns that link food processing levels to environmental impact. The Eco-Score, which evaluates factors such as carbon footprint, water usage, and biodiversity, provides a comprehensive measure of the environmental implications associated with different levels of food processing. NOVA 1 shows a fairly balanced distribution across Eco-Score grades, leaning slightly toward higher grades (A and B), which suggests lower environmental impact. Despite this tendency, 16.73% and 18.65% of products in NOVA 1 fall into lower grades D and E, accounting for 9,020 and 10,051 items, respectively, out of 53,888 items. NOVA 2, the class with the lowest count across all Eco-Score grades, still has 23.65% of its products in grade D and 4.64% in grade E, translating to 11,772 and 2,312 items, respectively, out of a total of 49,755 items. NOVA 3 presents a more even spread across Eco-Score grades, with 23.36% of products in grade D and 8.99% in grade E, equating to 23,250 and 8,950 items, respectively, out of 99,519 items. NOVA 4, however, stands out with the highest overall item count and a marked tendency toward lower Eco-Score grades, signifying greater environmental impact. In NOVA 4, 25.66% of products fall into grade C and 9.36% into grade D, representing 50,302 and 19,874 items out of 212,340 items. This trend underscores the association between higher processing levels and lower environmental grades.

The Chi-Square statistic varies among features, with the 'Category' feature having the highest value (1,417,026.89), indicating a strong overall association with NOVA groups, supported by a Cramer's V of 0.7678, reflecting a strong effect size. The 'Nutri-Score Grade' follows with a moderate effect size (Cramer's V = 0.2894). In contrast, 'Allergen,' 'Countries,' and 'Eco-Score Grade' show weaker associations, with Cramer's V values between 0.1120 and 0.1339, indicating small effect sizes. Refer to Supplementary Table 2 for the results of the Chi-Square test and Cramer's V.

The three box plots illustrate distinct patterns in Nutri-Score, Carbon Footprint, and Eco-Score across NOVA classes, with all features showing statistically significant associations with NOVA groups ($p < 0.05$). The Nutri-Score plot indicates increasing scores from NOVA 1 to 4, where NOVA 4 (ultra-processed foods) has the highest median and widest range, reflecting poorer nutritional quality. This is supported by a substantial Kruskal-Wallis statistic of 127986.31 and a moderate positive correlation of 0.40. The Carbon Footprint also rises across NOVA classes, with NOVA 4 exhibiting the highest impact, indicated by a Kruskal-Wallis statistic of 22.01 and a weak positive correlation of 0.12. In contrast, the Eco-Score shows less pronounced trends, with a slight decrease from NOVA 1 to 4 and a very weak negative correlation of -0.06, suggesting more processed foods may have lower Eco-scores. Refer to Supplementary Table 3 for results of the Kruskal-Wallis Test.

### 3.2 Predicting NOVA class with ML models

Table 1 presents the performance metrics of three ML models (LGBM, RF, and CB) across three different nutrient input scenarios (7, 8, and 44 nutrients). For the 7 and 8 nutrient scenarios, LGBM and RF show similar top performance, with LGBM slightly edging out in accuracy (0.84-0.85) and MCC (0.69). CB consistently trails behind, though it improves slightly with 8 nutrients. Interestingly, when using 44 nutrients, all models show a slight decrease in performance due to imputation, with RF maintaining the highest MCC (0.63) while LGBM and RF were tied with an accuracy of 0.80 and F1-Score of 0.79. When we used AEC to impute missing values and applied it in building ML models, the resulting accuracies ranged from 65% to 69%. For detailed model performance after AEC imputation, refer to Supplementary Table 4. These results were notably inferior compared to imputing with mean values.

The performance comparison across 10 folds for the best performing model shows that while the model achieves high accuracy and F1-Scores on the training data, there is a slight performance gap when compared to the validation set, suggesting a minor degree of overfitting. This gap, however, remains relatively negligible, reflecting the consistent application of the model's learned patterns across different folds. Despite extensive parameter tuning to control model complexity, LGBM, being a gradient boosting method, is inherently prone to overfitting due to its ability to capture intricate patterns within data. Nonetheless, the validation metrics are stable and align closely with test set performance, demonstrating the model's reliable generalization capability across this large and complex dataset.

Figure 7 shows SHAP summary plots that reveal the impact of various nutritional features on classifying foods into different NOVA classes, ranging from minimally processed (NOVA 1) to ultra-processed (NOVA 4). Each plot displays feature importance through violin shapes, where wider sections indicate a stronger influence on model predictions. For NOVA 1, 'Sodium' and 'Total Fat' are the most influential, suggesting their lower levels in MPFs.. In NOVA 2, 'Total Fat' and 'Energy' are key predictors, while 'Sugars total' and 'Carbohydrate' dominate for NOVA 3, highlighting their relevance in processed foods. NOVA 4, which represents ultra-processed items, is primarily influenced by high levels of 'Sugars total' and 'Sodium'. While folic acid and water content were prominent features in both Menichetti et al.'s and Arora et al.'s studies, our current analysis emphasizes other key nutrients like sugars total, sodium, and total fat across all NOVA categories (Arora et al., 2025; Menichetti et al., 2023). This strategic reduction in features aligns with our goal of creating a more streamlined classification model while maintaining discriminatory power through the most influential nutritional markers. This analysis provides insights into how specific nutritional components contribute to the classification of foods based on their processing level.

### 3.3 Validation of Models developed using FNDDS dataset

Upon validating the models built with various nutrient panels using the 2,970 food products from FNDDS 2009-10 data and their corresponding NOVA labels from Steele et al.(*FNDDS DOWNLOAD DATABASES : USDA ARS*, n.d.; Monteiro et al., 2019b), we achieved accuracy scores of 81.2%, 80.7%, and 77.8% on the best 7-nutrient, 8-nutrient, and 44-nutrient models, respectively. These results demonstrate that the model's performance on unseen data aligns closely with its testing performance, highlighting its strong generalizability and robustness.

## 4. Discussion

The application of machine learning to predict NOVA processing level using the Open Food Facts dataset shows promising results, specifically in the accurate classification of food products based on nutrient features using ensemble models like LGBM, RF, and CB. Notably, LGBM achieved accuracy in the 80% to 85% range for all three nutrient features despite the large data size. The best performance was achieved on the 8-nutrient model with an accuracy of 85%. This highlights the effectiveness of nutrient-based ML models in capturing the complexity of NOVA classification, particularly in distinguishing between minimally processed and ultra-processed foods.

Furthermore, exploratory analysis revealed several meaningful associations between NOVA classes and other attributes. A positive correlation was found between NOVA classification and Nutri-Score and carbon footprint scores, highlighting that higher food processing levels often correspond to lower nutritional quality and higher environmental impact. The study also suggests that UPFs tended to have lower Eco-Scores than MPFs, highlighting the environmental cost of heavy processing. The study found significant associations between NOVA class, food categories, and countries. Food categories like Cakes, Snacks, and Appetizers appeared more frequently in higher NOVA classes, showing that these food categories are more processed. The analysis of allergens across NOVA classes showed that gluten and milk are particularly present in UPFs. While analyzing the number of additives, they showed a strong positive correlation with NOVA classifications, where UPFs contained significantly more additives than MPFs. Through this analysis, we provide a detailed explanation of factors linked to food processing levels and illustrate how NOVA classes are related to nutritional quality, environmental impact, allergen presence, and additive use across different food categories and regions.

One of the shortcomings with the present dataset is that OFF is an open-source dataset, which means that the accuracy of the data and completeness can vary significantly. Information on some products may be incomplete and inconsistently updated, which could impact the reliability of the classifications. Additionally, cultural and regional biases may affect the distribution of food products in the dataset, which could mean that the models developed in the work may not be fully generalizable across different countries. Furthermore, imputing data can introduce biases or obscure relationships in the data.

Future research should aim to build a dataset with a global representation of food products and use that to automate classifications to improve generalizability. Secondly, investigating the impact of specific food additives and processing techniques on NOVA classifications could guide more nuanced classification systems that emphasize specific ingredients rather than relying on NOVA's broad categories. Thirdly, future research could delve into socio-economic factors like income level, cultural preferences, etc, and how they influence the consumption of UPFs. Finally, exploring the role of sustainability and environmental impact on food processing could extend the understanding of NOVA classifications from a health-centric perspective to one that includes environmental factors.

The findings of this study have significant applications in the food industry and for consumers. Manufacturers can utilize nutrient-based ML models to improve product formulations, reduce additives, and enhance the nutritional quality of their offerings. Retailers can leverage NOVA classifications to promote healthier food choices, aligning with growing consumer demand for transparency and health-conscious products. Policymakers can design targeted interventions to reduce the consumption of ultra-processed foods, addressing public health risks. For consumers, the deployment of a web server based on the developed ML model enables them to assess the processing level of food products directly by inputting product nutrient information, receiving instant NOVA classifications, and making informed choices about the nutritional and environmental impact of their food. This enables individuals to make more informed food choices by prioritizing minimally processed foods and developing healthier eating habits. By integrating this tool with industry and policy initiatives, it supports a comprehensive approach to developing healthier, more sustainable dietary practices.

## 5. Funding Source

The current work has not received any specific grant from any funding agencies.

## 6. Conflict of interest

Ganesh Bagler is the founder of Foodoscope Technologies Pvt. Ltd., a company working on computational gastronomy. This affiliation is disclosed in the interest of transparency. The research reported here was conducted without external financial support or influence from Foodoscope.

## 7. Authors' contributions

GB conceived the idea, designed the methodology, and supervised the project. NA collected the data, conducted the experiments, developed the models, and performed the literature survey. AC, MA, and SR implemented the webserver. SB, AD, AK, AS and AV deployed the server. NA and GB drafted the manuscript, proofread and approved the final manuscript.

## 8. Acknowledgments

GB thanks Indraprastha Institute of Information Technology Delhi (IIIT-Delhi) for the computational support. GB thanks Technology Innovation Hub (TiH) Anubhuti for the research grant. NA is an undergraduate research intern in the Complex Systems Laboratory and is thankful for the lab resources. This study was supported by the Infosys Center for Artificial Intelligence and Centre of Excellence in Healthcare at IIIT-Delhi.

## 9. Code Availability Statement

The underlying code for this study is available in GitHub and can be accessed via this [link](#).

## 10. Data Availability Statement

The datasets used and/or analysed during the current study are available in [Open Food Facts Database](#).

## 11. References

1. Monteiro, C. A., Cannon, G., Lawrence, M., Laura Da Costa Louzada, M. & Machado, P. P. *Ultra-Processed Foods, Diet Quality, and Health Using the NOVA Classification System Prepared By*.
2. Monteiro, C. A. *et al.* Ultra-processed foods: What they are and how to identify them. *Public Health Nutrition* vol. 22 936–941 Preprint at https://doi.org/10.1017/S1368980018003762 (2019).
3. Gibney, M. J. *Ultra-Processed Foods: Definitions and Policy Issues*. https://academic.oup.com/cdn/. (2018).
4. Popkin, B. M. Global nutrition dynamics: the world is shifting rapidly toward a diet linked with noncommunicable diseases. *Am J Clin Nutr* 84, 289–298 (2006).
5. Santos, F. S. Dos, Dias, M. da S., Mintem, G. C., Oliveira, I. O. de & Gigante, D. P. Food processing and cardiometabolic risk factors: a systematic review. *Rev Saude Publica* 54, 70 (2020).

6. Chen, X. *et al.* Consumption of ultra-processed foods and health outcomes: A systematic review of epidemiological studies. *Nutrition Journal* vol. 19 Preprint at https://doi.org/10.1186/s12937-020-00604-1 (2020).
7. Qu, Y. *et al.* Ultra-processed food consumption and risk of cardiovascular events: a systematic review and dose-response meta-analysis. *EClinicalMedicine* 69, (2024).
8. Isaksen, I. M. & Dankel, S. N. Ultra-processed food consumption and cancer risk: A systematic review and meta-analysis. *Clinical Nutrition* 42, 919–928 (2023).
9. Xiao, B. *et al.* Ultra-processed food consumption and the risk of incident chronic kidney disease: a systematic review and meta-analysis of cohort studies. *Renal Failure* vol. 46 Preprint at https://doi.org/10.1080/0886022X.2024.2306224 (2024).
10. Hecht, E. M. *et al.* Cross-sectional examination of ultra-processed food consumption and adverse mental health symptoms. *Public Health Nutr* 25, 3225–3234 (2022).
11. Lane, M. M. *et al.* Ultra-processed food exposure and adverse health outcomes: umbrella review of epidemiological meta-analyses. *BMJ* 384, (2024).
12. Mertens, E., Colizzi, C. & Peñalvo, J. L. Ultra-processed food consumption in adults across Europe. *Eur J Nutr* 61, 1521–1539 (2022).
13. Egnell, M. *et al.* Impact of the Nutri-Score front-of-pack nutrition label on purchasing intentions of individuals with chronic diseases: results of a randomised trial. *BMJ Open* 12, e058139 (2022).
14. Aceves-Martins, M. *et al.* Nutritional Quality, Environmental Impact and Cost of Ultra-Processed Foods: A UK Food-Based Analysis. *Int J Environ Res Public Health* 19, 3191 (2022).
15. Taillie, L. S. *et al.* The impact of an eco-score label on US consumers' perceptions of environmental sustainability and intentions to purchase food: A randomized experiment. *PLoS One* 19, e0306123 (2024).
16. Katidi, A. *et al.* Food Allergens in Ultra-Processed Foods According to the NOVA Classification System: A Greek Branded Food Level Analysis. *Nutrients* 15, 2767 (2023).
17. Amaraggi, B. *et al.* Ultra-processed food staples dominate mainstream U.S. supermarkets. Americans more than Europeans forced to choose between health and cost. *medRxiv* 2024.02.16.24302894 (2024) doi:10.1101/2024.02.16.24302894.
18. Messina, M. & Venter, C. Recent Surveys on Food Allergy Prevalence. *Nutr Today* 55, 22–29 (2020).
19. Allergens in food: scientific advice updated | EFSA. https://www.efsa.europa.eu/en/press/news/141126.
20. Eicher-Miller, H. A., Fulgoni, V. L. & Keast, D. R. Contributions of processed foods to dietary intake in the us from 2003-2008: A report of the food and nutrition science solutions joint task force of the academy of nutrition and dietetics, American society for nutrition, institute of food technologists, and international food information council. *Journal of Nutrition* 142, 2065–2072 (2012).
21. Chajès, V. *et al.* Ecological-Level Associations Between Highly Processed Food Intakes and Plasma Phospholipid Elaidic Acid Concentrations: Results From a Cross-Sectional Study Within the European Prospective Investigation Into Cancer and Nutrition (EPIC). *Nutr Cancer* 63, 1235–1250 (2011).
22. Moubarac, J.-C., Parra, D. C., Cannon, G. & Monteiro, C. A. Food Classification Systems Based on Food Processing: Significance and Implications for Policies and Actions: A Systematic Literature Review and Assessment. *Curr Obes Rep* 3, 256–272 (2014).
23. Asfaw, A. Does consumption of processed foods explain disparities in the body weight of individuals? The case of Guatemala. *Health Econ* 20, 184–195 (2011).

24. Monteiro, C. A., Cannon, G., Lawrence, M., Laura Da Costa Louzada, M. & Machado, P. P. *Ultra-Processed Foods, Diet Quality, and Health Using the NOVA Classification System Prepared By*.
25. Poti, J. M., Mendez, M. A., Ng, S. W. & Popkin, B. M. Is the degree of food processing and convenience linked with the nutritional quality of foods purchased by US households? *American Journal of Clinical Nutrition* 101, 1251–1262 (2015).
26. Bleiweiss-Sande, R. *et al.* Robustness of food processing classification systems. *Nutrients* 11, (2019).
27. Braesco, V. *et al.* Ultra-processed foods: how functional is the NOVA system? *Eur J Clin Nutr* 76, 1245–1253 (2022).
28. Menichetti, G., Ravandi, B., Mozaffarian, D. & Barabási, A. L. Machine learning prediction of the degree of food processing. *Nat Commun* 14, (2023).
29. Hu, G., Flexner, N., Tiscornia, M. V. & L'Abbé, M. R. Accelerating the Classification of NOVA Food Processing Levels Using a Fine-Tuned Language Model: A Multi-Country Study. *Nutrients 2023, Vol. 15, Page 4167* 15, 4167 (2023).
30. Arora, N., Bhagat, S., Dhama, R., & Bagler, G. (2025). Machine learning and natural language processing models to predict the extent of food processing. Journal of Food Composition and Analysis, 146, 107938. https://doi.org/10.1016/j.jfca.2025.107938.
31. Monteiro, C. A. *et al.* Ultra-processed foods: What they are and how to identify them. *Public Health Nutrition* vol. 22 936–941 Preprint at https://doi.org/10.1017/S1368980018003762 (2019).
32. FNDDS DOWNLOAD DATABASES : USDA ARS. https://www.ars.usda.gov/northeast-area/beltsville-md-bhnrc/beltsville-human-nutrition-research-center/food-surveys-research-group/docs/fndds-download-databases/.
33. https://in.openfoodfacts.org/.
34. Pereira, R. C., Santos, M. S., Rodrigues, P. P. & Abreu, P. H. Reviewing Autoencoders for Missing Data Imputation: Technical Trends, Applications and Outcomes. *Journal of Artificial Intelligence Research* 69, 1255–1285 (2020).
35. Menichetti, G. & Barabási, A. L. Nutrient concentrations in food display universal behaviour. *Nature Food 2022 3:5* 3, 375–382 (2022).
36. Krzywinski, M. & Altman, N. Visualizing samples with box plots. *Nat Methods* 11, 119–120 (2014).
37. Yun, J. & Kim, H. ZNorm: Z-Score Gradient Normalization for Deep Neural Networks. (2024).
38. Akiba, T., Sano, S., Yanase, T., Ohta, T. & Koyama, M. Optuna: A Next-generation Hyperparameter Optimization Framework. *Proceedings of the ACM SIGKDD International Conference on Knowledge Discovery and Data Mining* 2623–2631 (2019) doi:10.1145/3292500.3330701.
39. Breiman, L. *Random Forests*. vol. 45 (2001).
40. Ke, G. *et al. LightGBM: A Highly Efficient Gradient Boosting Decision Tree*. https://github.com/Microsoft/LightGBM.
41. Prokhorenkova, L., Gusev, G., Vorobev, A., Dorogush, A. V. & Gulin, A. CatBoost: unbiased boosting with categorical features. *Adv Neural Inf Process Syst* 2018-December, 6638–6648 (2017).
42. Bergstra, J., Ca, J. B. & Ca, Y. B. *Random Search for Hyper-Parameter Optimization Yoshua Bengio*. *Journal of Machine Learning Research* vol. 13 http://scikit-learn.sourceforge.net. (2012).

43. Chawla, N. V, Bowyer, K. W., Hall, L. O. & Kegelmeyer, W. P. *SMOTE: Synthetic Minority Over-Sampling Technique*. *Journal of Artificial Intelligence Research* vol. 16 (2002).
44. Prusty, S., Patnaik, S. & Dash, S. K. SKCV: Stratified K-fold cross-validation on ML classifiers for predicting cervical cancer. *Frontiers in Nanotechnology* 4, (2022).
45. Lundberg, S. & Lee, S.-I. A Unified Approach to Interpreting Model Predictions. (2017).

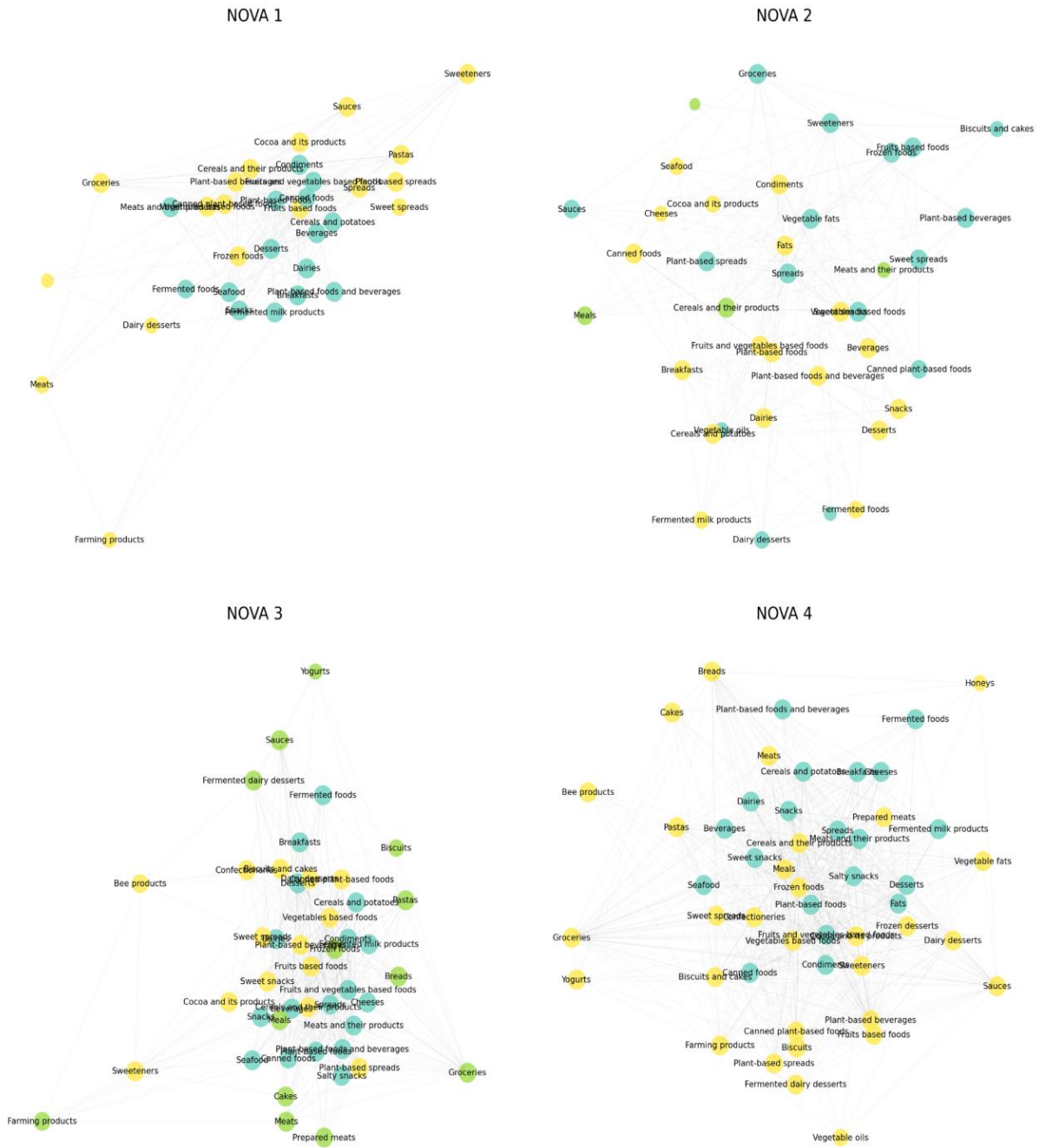

**Figure 1: Network of Top 50 categories across the four NOVA classes**. Each node stands for a category, with yellow nodes representing parent categories, green nodes indicating child categories, and blue nodes showing sub-child categories. The node size represents the degree, the number of other categories it is related to. The edge represents connection between every pair of nodes. The categories are hierarchically structured as X:Y:Z, where X represents the parent category, Y the child category, and Z the sub-child category. This hierarchical organization allows for a clearer visualization of relationships and groupings across different NOVA classes, revealing patterns and connections unique to each class.

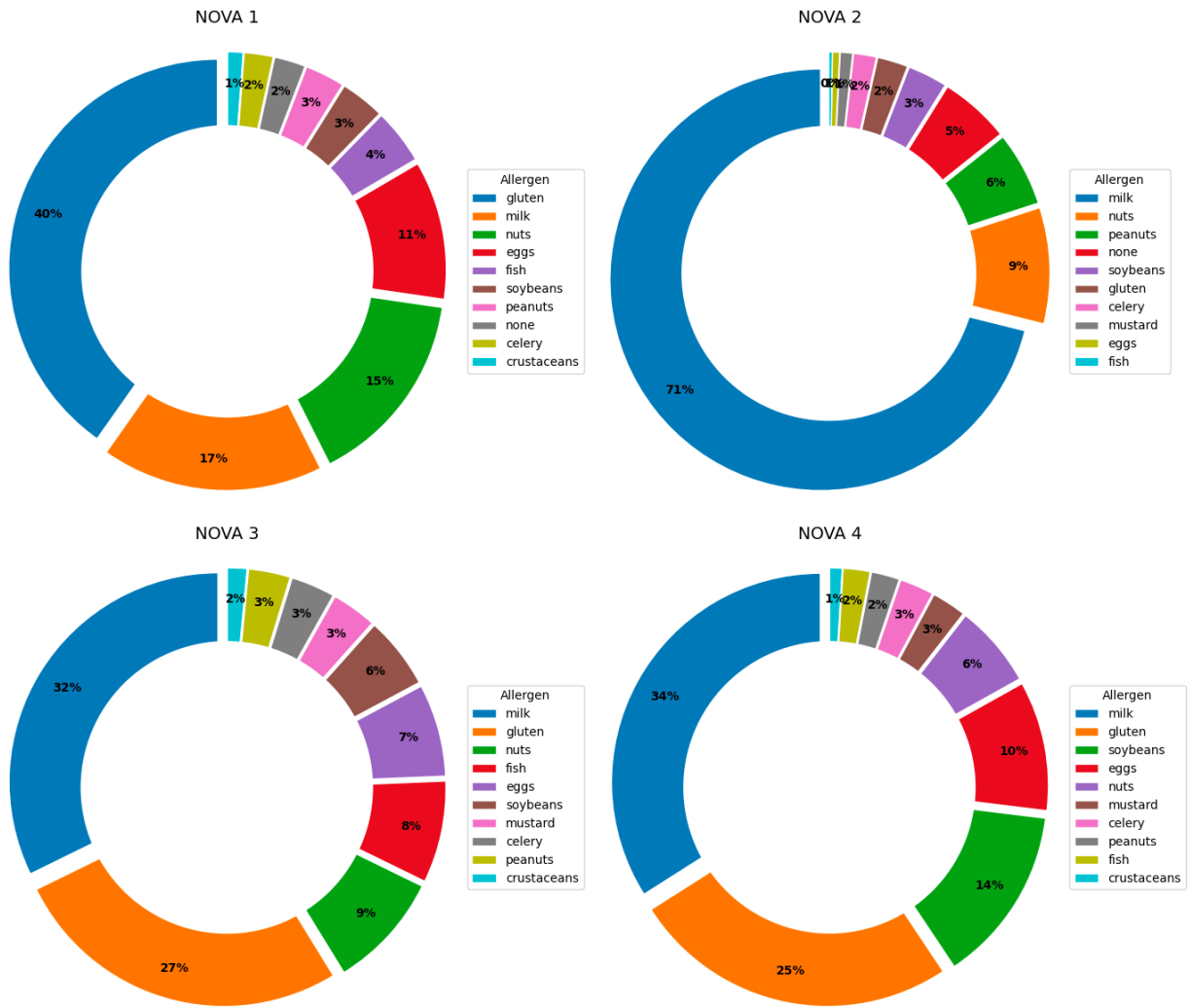

**Figure 2: Presence of allergens across the four NOVA classes.** The statistics reflect the percentage distribution of different allergens within each NOVA class. Each segment in the donut chart corresponds to a specific allergen category, with the size of the segment representing the proportion of foods containing that allergen.

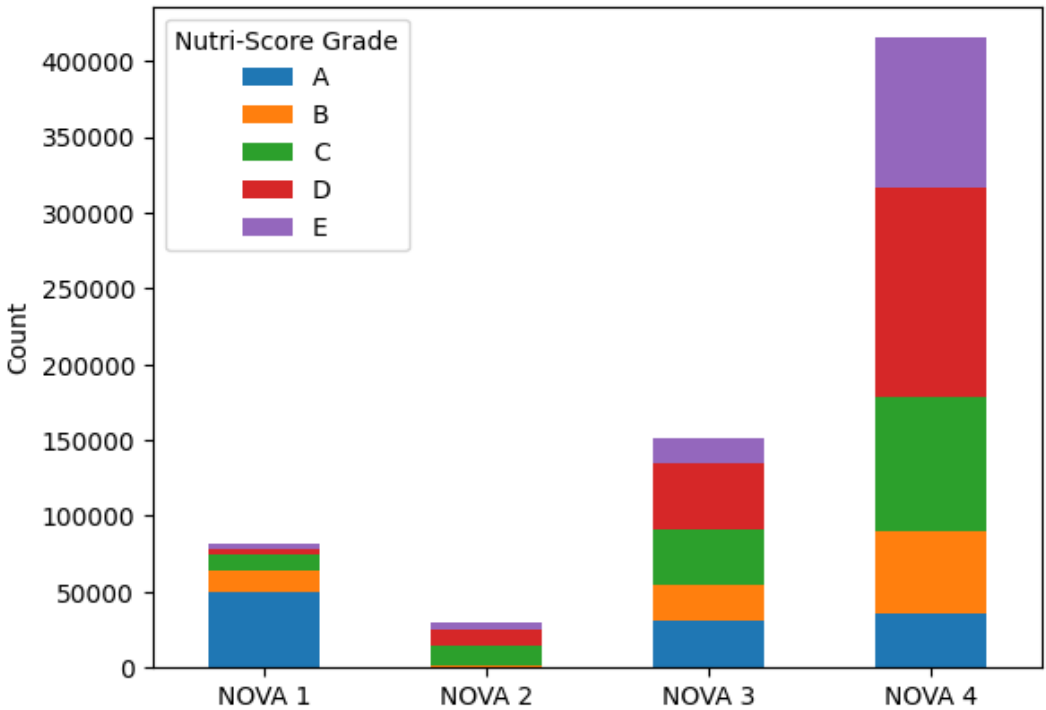

**Figure 3: The visualization presents the distribution of Nutri-Score grades (A through E) across the four NOVA classes.** This stacked bar chart demonstrates how nutritional quality varies across different levels of food processing.

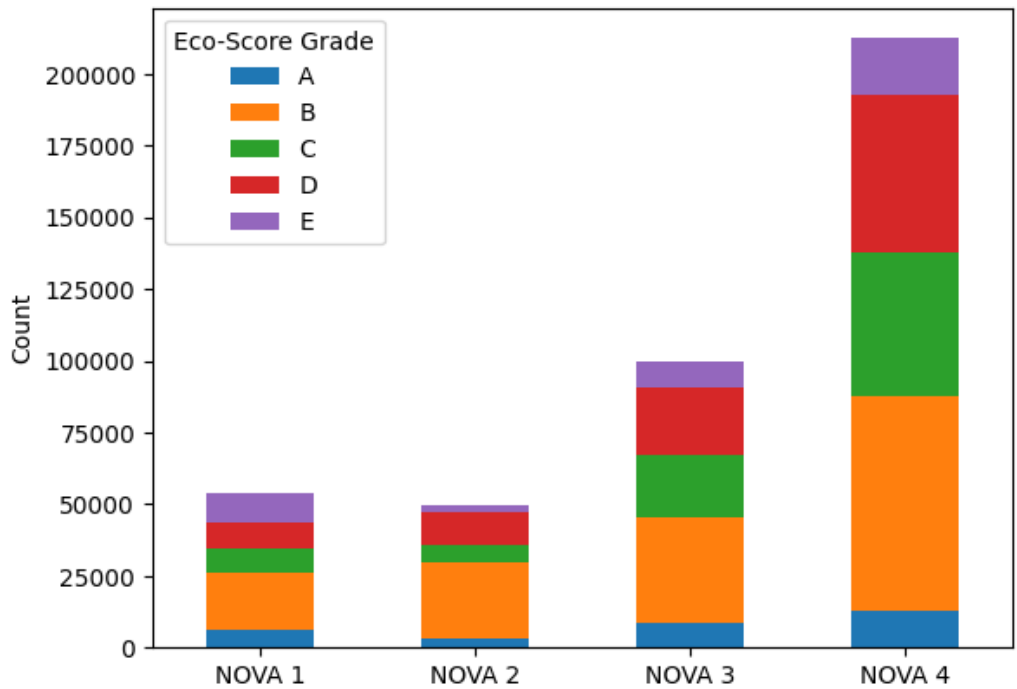

**Figure 4: The visualization presents the distribution of Eco-Score grades (A through E) across the four NOVA classes.** This stacked bar chart demonstrates how environmental quality varies across different levels of food processing.

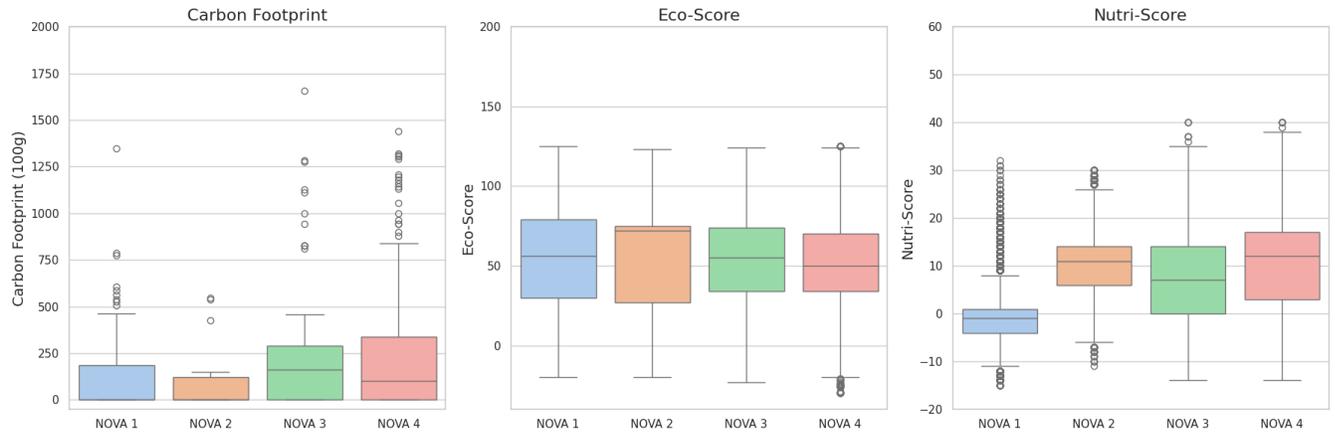

**Figure 5: Box plots illustrating the relationships between three key nutritional and environmental metrics across NOVA classes.** The leftmost plot reveals carbon footprint measurements, the center plot displays Eco-Score distributions and the rightmost plot presents Nutri-Score distributions.

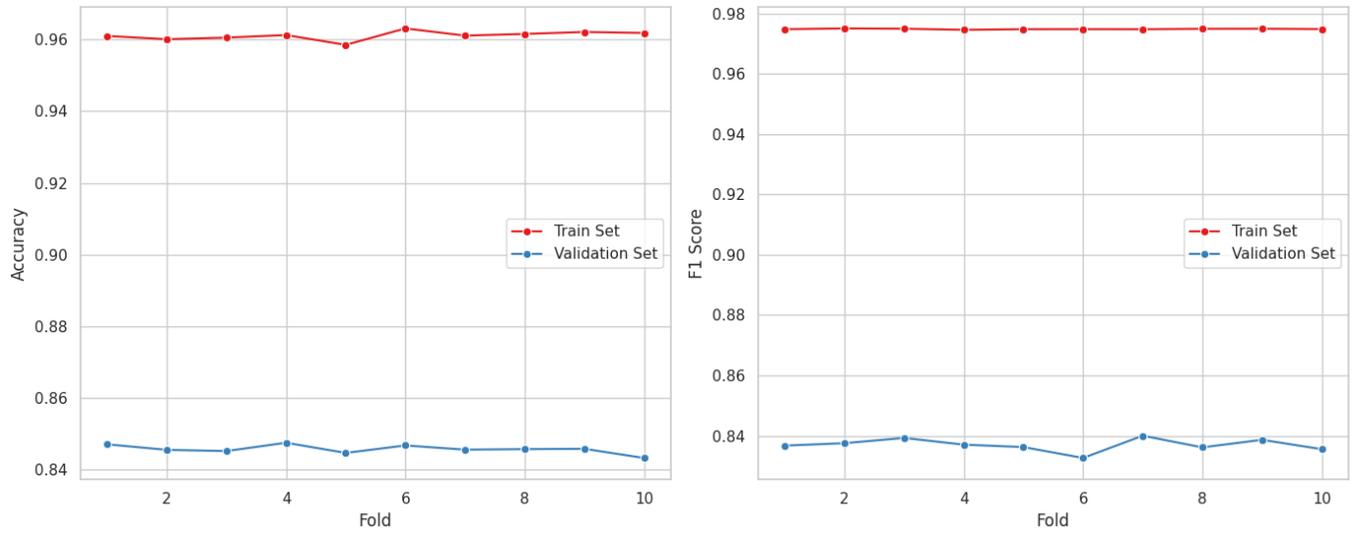

**Figure 6: Cross-validation performance metrics for the Light Gradient Boosting Machine (LGBM) model trained on eight key nutritional parameters.** The dual-line plots demonstrate the model's stability and generalization capability across ten distinct validation folds. The red line shows the model's performance on the train set while the blue line shows the model's performance on the validation set. The left plot shows Accuracy, while the right shows F1-score performance.

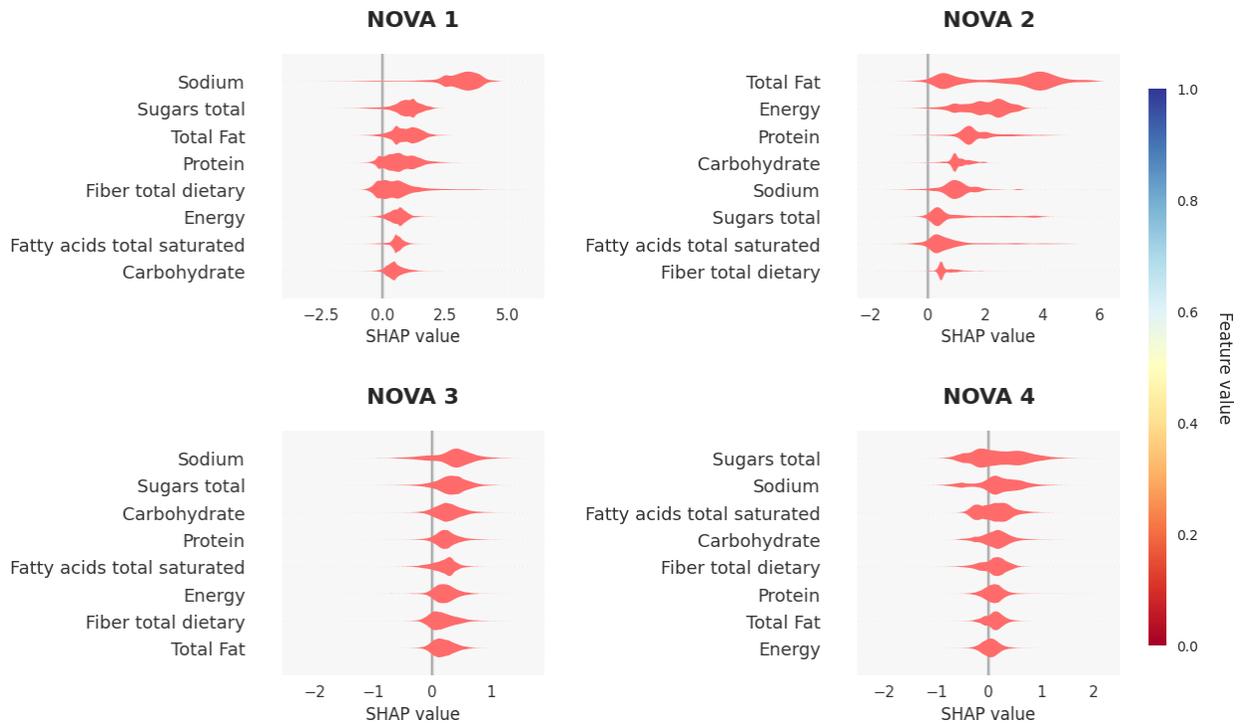

**Figure 7: SHAP value distributions for the eight key nutritional parameters across all four NOVA classes, illustrating feature importance and their impact on model predictions.** The violin plots demonstrate the magnitude and direction of each nutrient's contribution to the classification decisions. The width of each violin shape indicates the density of SHAP values at different points, while the position relative to zero shows whether the feature contributes positively or negatively to the classification. The color gradient from blue to red reflects the feature values, with higher values shown in darker red.

| Metric | CB | LGBM | RF |
|--------|------|------|------|
| 7 Nutrients | | | |
| ACC | 0.8 | **0.84** | 0.84 |
| F1 | 0.79 | **0.84** | 0.83 |
| MCC | 0.61 | **0.69** | 0.68 |
| 8 Nutrients | | | |
| ACC | 0.81 | **0.85** | 0.84 |
| F1 | 0.8 | **0.84** | 0.83 |
| MCC | 0.63 | **0.69** | 0.68 |
| 44 Nutrients | | | |
| ACC | 0.78 | 0.8 | **0.8** |
| F1 | 0.76 | 0.79 | **0.79** |
| MCC | 0.58 | 0.62 | **0.63** |

**Table 1: Model performance across 7, 8 and 44 nutrients.** For 44 nutrient feature space, we used mean values to fill the missing values, while for 7 and 8 nutrient feature space we have dropped all entries with missing values.

# SUPPLEMENTARY INFORMATION

**Table S1: Number and percentage of food products with missing values**

| Nutrient | Total Food Products with None | Percentage of Total Food Products |
|---|---|---|
| 6:00 | 875065 | 0.999988572 |
| 22:01 | 875065 | 0.999988572 |
| 8:00 | 875061 | 0.999984001 |
| 10:00 | 875058 | 0.999980573 |
| 14:00 | 875051 | 0.999972574 |
| 18:00 | 875045 | 0.999965717 |
| 4:00 | 875044 | 0.999964574 |
| 12:00 | 875044 | 0.999964574 |
| 16:00 | 875023 | 0.999940577 |
| Carotene, beta | 875001 | 0.999915436 |
| 18:01 | 874958 | 0.999866297 |
| Choline, total | 874956 | 0.999864012 |
| 20:5 n-3 | 874895 | 0.999794303 |
| 20:04 | 874872 | 0.99976802 |
| 22:6 n-3 | 874779 | 0.999661743 |
| 18:02 | 874583 | 0.999437762 |
| Caffeine | 874240 | 0.999045796 |
| 18:03 | 874078 | 0.998860669 |
| Vitamin K (phylloquinone) | 873492 | 0.998191012 |
| Selenium | 872398 | 0.996940834 |
| Copper | 870834 | 0.995153558 |
| Folate, total | 866847 | 0.990597377 |
| Zinc | 864583 | 0.988010171 |
| Phosphorus | 861110 | 0.984041368 |
| Vitamin B-12 | 860623 | 0.983484844 |
| Magnesium | 860396 | 0.983225438 |
| Vitamin D (D2 + D3) | 860286 | 0.983099734 |
| Vitamin B-6 | 858246 | 0.980768506 |
| Alcohol | 855201 | 0.977288804 |
| Fatty acids, total polyunsaturated | 826418 | 0.944396766 |
| Fatty acids, total monounsaturated | 826375 | 0.944347627 |
| Potassium | 780338 | 0.891738422 |
| Vitamin C | 666614 | 0.761779276 |
| Iron | 617469 | 0.705618376 |
| Cholesterol | 613224 | 0.700767363 |
| Calcium | 609769 | 0.69681913 |
| Fiber, total dietary | 358427 | 0.409595749 |
| Energy | 131423 | 0.150184841 |
| Fatty acids, total saturated | 130966 | 0.1496626 |
| Sugars, total | 108892 | 0.124437334 |
| Sodium | 108589 | 0.124091078 |
| Protein | 94214 | 0.107663915 |
| Carbohydrate | 93386 | 0.10671771 |
| Total Fat | 93067 | 0.10635317 |

Table S2: Relationship between NOVA class and other features using Chi-Square Test and Cramers V

| Feature | Chi-Square Test | | Cramers V |
|---|---|---|---|
| | Chi Square Statistic | P-Value | |
| Category | 1417026.89 | 0 | 0.7678 |
| Nutriscore Grade | 171576.43 | 0 | 0.2894 |
| Allergen | 23442.03 | 0 | 0.1339 |
| Ecoscore Grade | 15825.18 | 0 | 0.112 |

| Table S3: Relationship between NOVA class and other features using Kruskal Wallis and Correlation |||||
|---|---|---|---|
| | **Kruskal Wallis Test** | | |
| **Feature** | **Kruskal Wallis Statistic** | **P-Value** | **Correlation** |
| Number of Additives | 375971.8 | 0 | 0.42 |
| NutriScore | 127986.31 | 0 | 0.4 |
| Carbon Footprint | 22.01 | 6.48E-05 | 0.12 |
| Ecoscore | 3929.469 | 0 | -0.06 |

| Table S4: Model performance using Autoencoders to fill missing values in 44 nutrient data | | | |
|---|---|---|---|
| **Model** | **CB** | **RF** | **LGBM** |
| ACC | 0.64 | 0.67 | **0.68** |
| F1-Score | 0.52 | 0.69 | **0.69** |
| MCC | 0.2 | 0.33 | **0.33** |